\documentclass{WileyMSP-template}
\usepackage{comment}
\usepackage{amsmath}
\usepackage{amssymb}
\usepackage{physics}
\newcommand{\iu}{\mathrm{i}\mkern1mu}
 
\usepackage{xcolor} 
\usepackage{comment}

\usepackage{multirow}

\begin{document}

\pagestyle{fancy}
\rhead{\includegraphics[width=2.5cm]{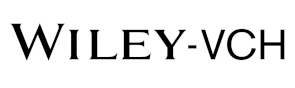}}

\title{Near-infrared polarimetric imaging with nonlinear flat-optics  \\
}

\maketitle


\author{Evgenii Menshikov, Marco A. López Sánchez, Paolo Franceschini,  Andrea Tognazzi, Domenico de Ceglia, Kristina Frizyuk, Costantino De Angelis$^*$},



\begin{affiliations}
Evgenii Menshikov \\
Università degli studi di Brescia, Via Branze 38, Brescia, Italy\\

Marco A. López Sánchez\\
Photonics Research Group, Ghent
University-IMEC, Technologiepark-Zwijnaarde 123, Ghent, Belgium\\

Dr. Kristina Frizyuk \\ Karlsruhe Institute of Technology, Kaiserstrasse, 12, Karlsruhe, 76131, Germany

Dr. A. Tognazzi\\
Università degli studi di Palermo, Viale delle Scienze ed. 10, Palermo, 90128, Italy\\
Istituto Nazionale di Ottica - Consiglio Nazionale delle Ricerche (INO-CNR), Via Branze 45, Brescia, Italy\\

Dr. P. Franceschini, Prof. Domenico de Ceglia, Prof. C. De Angelis\\
Università degli studi di Brescia, Via Branze 38, Brescia, Italy\\
Istituto Nazionale di Ottica - Consiglio Nazionale delle Ricerche (INO-CNR), Via Branze 45, Brescia, Italy\\
Email Address: costantino.deangelis@unibs.it\\

\end{affiliations}



\begin{abstract}
A compact and broadband polarimetric imaging platform is presented, based on second-harmonic generation (SHG) in nonlinear flat-optics. 
The system employs periodic all-dielectric AlGaAs gratings to induce polarization-dependent SH emission, enabling pixel by pixel direct retrieval of the full Stokes vector from an input intensity distribution in the near-infrared range.
By engineering the geometry and orientation of the polarimetric units, sensitivity to linear and circular polarization components is achieved. 
A superpixel design comprising four polarimetric structures allows accurate reconstruction of the polarization state without moving parts or sequential measurements. 
This approach offers a scalable, passive, and cost-effective solution for polarimetric imaging, particularly suited for near-infrared applications.
\end{abstract}


\section{Introduction}

An electromagnetic (EM) monochromatic plane wave 
has four degrees of freedom, namely the amplitude $A$, the phase $\phi$, the polarization orientation $\psi$, and ellipticity $\chi$~\cite{BornWolf}. 
In fact, the parameters $\psi$ and $\chi$ 
describe the so-called State of Polarization (SoP) and specify the properties of the electric field oscillations in the plane perpendicular to the propagation direction. 
This intrinsic property of radiation plays a crucial role in both the fundamental understanding of light features and in a variety of applications~\cite{rubin2022polarization}.  
Indeed, the measurements and the analysis of the SoP (\textit{i.e.}, polarimetry \cite{Tyo2006}) is of paramount importance in various fields, such as astronomy~\cite{Cotton2017}, remote sensing~\cite{Tyo2006}, quantum optics~\cite{Lodahl2017}, and biology~\cite{ghosh2011tissue}. 
In particular, imaging polarimetry (the technique aiming at mapping the SoP across a vast scene of interest), given its ability to analyze spatially varying light beams, allows one to retrieve information about the shape~\cite{garcia2015surface, morel2006active, atkinson2006recovery} and arrangement~\cite{malinowski2023polarimetric, shkuratov2007multispectral} of the reflecting structures, the orientation of light emitters~\cite{brasselet2023polarization, kim2021measuring}, or the optical activity of various materials~\cite{schellman1975circular}. 
Therefore, in the last decades, several methods have been developed to measure the SoP in an extended scene~\cite{Tyo2006}.\\


A widely spread representation of the SoP is given by the Stokes vector formalism~\cite{BornWolf,Schaefer2007}. 
Within this framework, the SoP of an arbitrary polarized wavefront is represented as a four-element vector, \textit{i.e.}, the Stokes vector, $\vb{S}=\left( S_0, \, S_1, \, S_2, \, S_3  \right)$. 
Given a Cartesian reference frame $xyz$, with $z$ being the propagation direction, the vector components are defined as $S_0=I$, $S_1=I_x-I_y$, $S_2=I_{L+45}-I_{L-45}$, and $S_3=I_\text{RCP}-I_\text{LCP}$. In the previous expressions, $I$ is the total intensity of the light-field and the terms $I_x$, $I_y$, $I_{L+45}$, and $I_{L-45}$ are the intensity of light in linear polarization component along the $x$ (horizontal, H), $y$ (vertical, V), $+45^\circ$, and $-45^\circ$ direction, respectively.
The terms $I_\text{RCP}$ and $I_\text{LCP}$ denote the intensity of the right-hand (RCP) and left-hand (LCP) circularly polarized light, respectively. From a conceptual point of view, the Stokes vector formalism allows to describe the SoP of the input light as the projection on three polarization basis sets: \textit{i)} H/V ($S_1$), \textit{ii)} $\pm45^\circ$ ($S_2$), and \textit{iii)} RCP/LCP ($S_3$). 
Practically, this means that the value of the components $S_0$, $S_1$, $S_2$, and $S_3$ can be directly determined by measuring the intensity (or power) in the different polarization bases. 
From an experimental point of view, this is a big advantage, since standard photodetectors operating at optical frequencies typically respond to changes in the intensity of the EM field (not to its phase).\\


A typical polarimetry assessment setup always contains two parts: the polarization analyzer and the detection system (electric readout). The light detector probes the output response (such as an optical signal or photocurrent) resulting from the interaction between the analyzer and the incident radiation with unknown SoP. 
In the traditional approach, the radiation under analysis propagates sequentially through rotating polarizing elements (optical analyzer) and the transmitted intensity is measured by a photodetector~\cite{Schaefer2007}. 
The SoP is thus determined from several (a minimum of four) intensity measurements by properly arranging the polarizing elements in front of the detector. 
This takes relatively long acquisition times and it is not capable to monitor transient events. Another method consists in performing parallel measurements by splitting the beam into several optical paths and using multiple polarizers and detectors~\cite{Tyo2006,Compain1998}, but result in inherently complex and bulky systems. These limitations were alleviated with the development of integrated polarimetric devices employing variable liquid crystal retarders~\cite{Wolff1997, DeMartino2003}. These devices are designed to dynamically modulate the retardance values under an external voltage, allowing the full polarization state to be reconstructed from sequential measurements. 
More compact implementations include thin-film analyzers (micropolarizer gratings)~\cite{Guo2000,Gruev:07} and devices based on two-dimensional (2D) materials, exploiting their anisotropic absorption~\cite{yuan2015polarization}. While both types of devices exhibit strong sensitivity to linear polarization, their response to circular components is usually absent or quite limited, requiring more advanced designs~\cite{fan2023broadband, bachman2012spiral,zhang2024advanced}. 

In the past decade, metasurfaces (optical elements composed of judiciously arranged nanoscatterers) have opened new avenues in the manipulation of light's amplitude, phase, and SoP~\cite{Yu2011,Yu2014,Lalanne1998}. 
Among a wide variety of applications, dielectric metasurfaces have been successfully employed also for imaging polarimetry~\cite{Arbabi2018,Rubin2019, intaravanne2020recent}. 
In these devices, the extended wavefront impinges onto the polarimetric unit composed of an ordered array of unit elements with specific shapes and/or orientations. 
The individual polarimetric units (called super-pixels) comprise a set of optical metasurfaces, each one being designed to respond only to one of the three polarization basis (\textit{e.g.}, H/V, $\pm45^\circ$, or RCP/LCP) and it splits the two orthogonal states of polarization to different points on the detection plane. 
The determination of the Stokes parameters is achieved by measuring the intensity at image sensor pixels on the detector plane, corresponding to one single super-pixel on the analyzer plane. 
Further advances along this path are represented by the integration of a metasurface analyzer directly with a photodetector, enabling an extremely compact and self-contained polarimetric platform~\cite{Li2020,Tian2022, zuo2023chip}.
In this case, the individual super-pixel comprises a few sensing units (three or four depending on the implementations), whose output photocurrent depends on the SoP of the input light. 
The Stokes parameters are extracted by measuring the photocurrent levels from the various sensing units, each one addressing one specific polarization basis. Successful implementations of this working principle, including, for example, integrated metasurface polarization filters~\cite{zuo2023chip}, chiral plasmonic metasurfaces integrated with graphene-silicon photodetectors~\cite{Li2020}, or Weyl semimetal sensing units accompanied by integrated grating waveplates~\cite{Tian2022}.

Here, we propose a novel method to determine the SoP of polarized light at fundamental frequency (FF) based on second-harmonic (SH) generation process in a nonlinear all-dielectric platform.
We show that in specifically optimized periodic structures, the intensity of SH radiation depends on the SoP of the FF, allowing for full Stokes polarization retrieval.
Namely, we show that polarization-dependent SH generation can be induced by grating units -- fabricated from $[001]||z$-oriented aluminum gallium arsenide (AlGaAs) thin film -- each featuring an identical, nonchiral geometry but differing in the bars orientation within the $\langle 001\rangle$ plane of the crystalline lattice.
To fully explain the process, we introduce an analytical description of the properties of the nonlinear radiation from the periodic polarimetric unit. 
Finally, we numerically demonstrate the capability of a four-unit super-pixel as polarization analyzer for imaging polarimetry. 
We show that the full set of Stokes parameters can be extracted from the intensity measurements of the SH radiation from individual scatterers using an image sensor placed at the detection plane.
Our results indicate that nonlinear optics holds strong potential for developing broadband imaging polarimetry devices with low fabrication complexity suitable for near-IR operation.

\begin{figure}[!t]
  \includegraphics[width=1\textwidth]{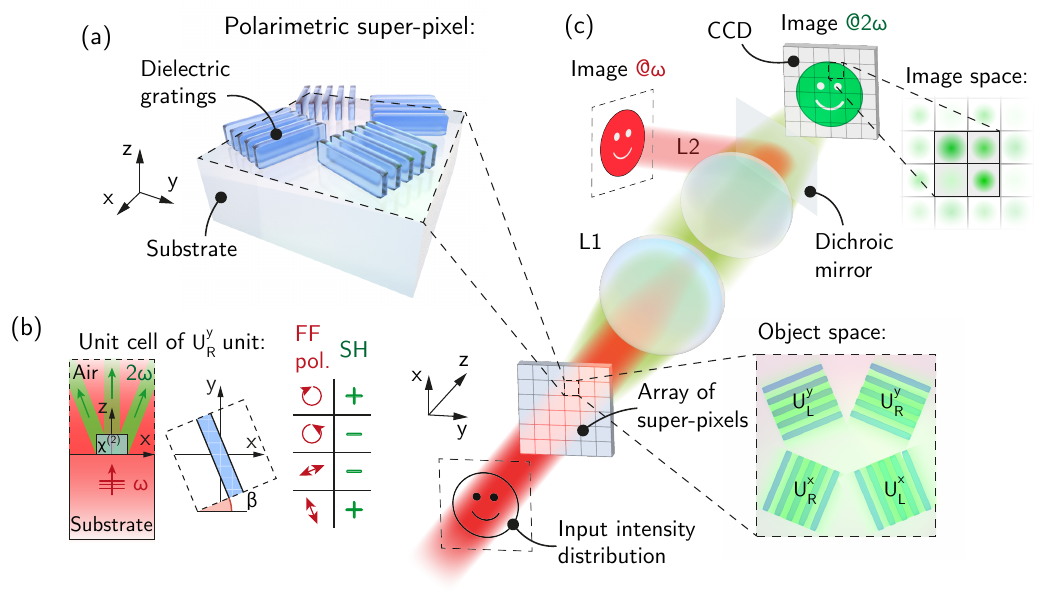}
  \caption{ Visualization of the nonlinear polarimetric device. 
  (a) Polarimetric super-pixel consists of four all-dielectric gratings tilted by $\pm\pi/8$ $(\pm22.5^\circ)$ with respect to the crystalline lattice of the AlGaAs film. 
  (b) Under illumination by a pump irradiation, the fundamental field generates second harmonic signal in the $\chi^{(2)}$ layer, which is most prominent for RCP or linear input parallel to the grating bars, and negligible for the LCP or LP orthogonal to the bars.
  (c) SH response from the polarimetric units arranged in an array of super-pixels is relayed by the telescope formed by lenses $L_1$ and $L_2$ on the CCD, allowing for the imaging polarimetry. }
  \label{fig:concept_vis}
\end{figure}

\section{Theory and Modeling}

Figure~\ref{fig:concept_vis} illustrates the concept of a nonlinear all-dielectric polarimetric imaging device. 
The device consists of an array of polarimetric super-pixels, comprising four gratings made of a thin layer of nonlinear medium (AlGaAs) on a $\text{Al}_2\text{O}_3$ substrate, as shown in Figure~\ref{fig:concept_vis}a. 
The gratings have the same geometrical sizes, but are rotated at different angles $\beta$ with respect to the crystalline lattice. Depending on the rotation angle, we denote them as $U^x_L$ and $U^x_R$ for the units rotated with respect to the $x$-axis, and $U^y_L$ and $U^y_R$ for those rotated with respect to the $y$-axis, while the $R$ and $L$ superscripts indicate the handedness that maximizes the SH signal.
The units are illuminated from the substrate side and generate SH in the open 0 and $\pm1$ diffraction orders in air (in the operating range 1450-1650 nm), as shown in the left panel of Figure~\ref{fig:concept_vis}b. Further, when referring to the SH signal we mean total contribution from all diffraction orders.
After nonlinear interaction, both the transmitted fundamental light ($\omega$) and the generated SH signal ($2\omega$) are collected with a lens (see Figure~\ref{fig:concept_vis}c). Next, the SH signal is relayed onto a CCD camera via a telescope system for analysis of the polarization dependent nonlinear response. The fundamental harmonic, which is only weakly influenced by its interaction with the units, can be redirected away from the main optical path using a dichroic mirror, which transmits the SH signal and reflects the pump.

Twist between the gratings and the lattice allowed us to achieve discrimination between the right and left-hand polarization states. Here we exploit the phenomenon of circular dichroism at second harmonic (SH-CD), or the sensitivity of the second-order nonlinear response to the handedness of the circular polarization \cite{CarlettiRocco2024}, which quantitatively can be expressed through the ratio
$\text{SH-CD} = (I^{2\omega}_\text{RCP} - I^{2\omega}_\text{LCP})/(I^{2\omega}_\text{RCP} + I^{2\omega}_\text{LCP})$~\cite{nikitina2024achiral}.
As we will show below, the symmetry of the nonlinear crystal and the structure give rise to particular angular terms in the dependence of SH power on the input polarization orientation angle, $\psi$, with their relative weights determined by the value of the SH-CD. 
We leverage this property to obtain strong SH contrast between RCP and LCP excitation and a distinct SH response for input fields oriented parallel or orthogonal to the grating bars, as schematically summarized in Figure~\ref{fig:concept_vis}b. Generally, the suggested units generate a SH response similar to the response of a rotated elliptical polarizer,
allowing for polarimetric measurements using a set of units with identical geometries oriented differently in space~\cite{hsu2014full}.

The geometry of the device was optimized in COMSOL Multiphysics software to exhibit maximal SH circular dichroism ($\text{SH-CD}=0.981$ with input at 1550 nm), resulting in a value of the SH linear dichroism of $\text{SH-LD} = (I^{2\omega}_{||} - I^{2\omega}_{\perp})/(I^{2\omega}_{||} + I^{2\omega}_{\perp})=0.956$ (see Suppl. Inf. 1). The corresponding geometrical parameters are a bar width $w = 653$ nm, a period $\Lambda = 1316$ nm, and a height $h = 246$ nm. We find that the simultaneous maximization of LP and CD in our design leads to minor variations in the optimized geometrical parameters (see Suppl. Inf. 2). 

\begin{figure}[!t]
\centering
  \includegraphics[width=0.9\textwidth]{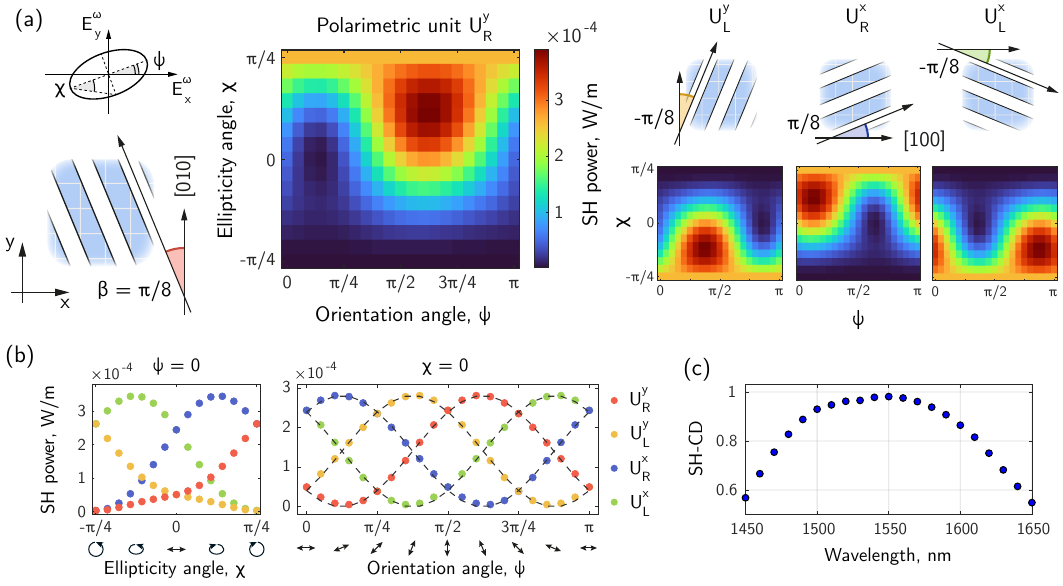}
  \caption{Second harmonic response from the polarimetric units with input intensity $I^{\omega}_0\!=\!1$~MW/cm$^2$. (a) SH map at 1550 nm for the $U^y_R$ unit, rotated by the angle $\beta=\pi/8$ with respect to [010] direction ($y$-axis), as a function of the ellipticity and orientation angles of the polarization ellipse. Structure $U^y_L$ ($\beta=-\pi/8$) features inverted map, showing the opposite dependence to the handedness of the input. 
  (b) Dependence of SH signal power on the orientation angle $\psi$ is similar to the response of a linear polarizer $I\propto\cos^2\psi$ depicted by dashed lines; solid lines correspond to slices of SH maps with $\chi=0$. 
  (c) Dependence of the nonlinear circular dichroism on the input wavelength features a maximum near 1550 nm.  }
  \label{fig:scatts_expl}
\end{figure}

\subsection{Nonlinear response of polarimetric units}
Figure \ref{fig:scatts_expl}a shows the numerical simulation of the SH power emitted by the polarimetric units following FF excitation, whose polarization state traverses the entire Poincar\'e sphere. Here we utilize the formalism of a polarization ellipse SoP which is parametrized by the ellipticity angle $\chi \in [{-\pi/4}, {\pi/4}]$ and orientation angle $\psi \in [0, \pi]$ (see inset in Figure~\ref{fig:scatts_expl}a). From the map of the polarimetric unit $U^y_R$, having a $\pi/8$ tilt with respect to the [010] direction, one can observe strong dependence of the nonlinear response on the input SoP. Rotation of the periodic structure by the same angle, but in the opposite direction compared to $U^y_R$, \textit{i.e.}, $\beta=-\pi/8$, results in a map obtained through inversion about the point $(\chi, \psi) = (0, \pi/2)$, corresponding to the behaviour of a $U^y_R$ unit. In turn, units $U^x_L$ and $U^x_R$, tilted with respect to the orthogonal [100] direction, demonstrate a $\pi/2$ shift along the $\psi$ axis, compared to the corresponding maps of $U^y_{R,L}$ units. Left panel of Figure~\ref{fig:scatts_expl}b shows the vertical slices of the maps ($\psi = 0$). Here one can see that $U^{x,y}_R$ units generate a negligibly small SH signal under LCP input ($\chi = -\pi/4$), with the signal increasing toward the RCP input ($\chi = \pi/4$). As expected, $U^{x,y}_L$ units feature the opposite dependence on the ellipticity angle. In case of linearly polarized input (Figure~\ref{fig:scatts_expl}b, right panel, $\chi = 0$), the dependencies on the orientation angle $\psi$ exhibit one maximum and one minimum for each value of the ellipticity angle, approximating shifted $\cos^2(\psi+\psi_0)$ dependence, showed by dashed lines. We observe that the SH-CD value monotonically decreases as the FF wavelength is shifted away from 1550 nm, at which it reaches its maximum, and drops to approximately 0.6 for a detuning of $\pm100$ nm (see Figure~\ref{fig:scatts_expl}c). 
The nonlinear conversion efficiency of the unit can be described in terms of the nonlinear conversion coefficient $\xi^{2\omega}=P^{2\omega}/(P^{\omega})^2$, where $P^{2\omega}$ and $P^{\omega}$ is the power of the output SH and  input FF radiation, respectively. From the numerical results shown in Figure~\ref{fig:scatts_expl}, we obtain $\xi^{2\omega}\simeq4 \times 10^{-9}$~W$^{-1}$ (see Suppl. Inf.~3.1 for more details). We also note that the linear transmittance of the units shows relatively weak dependence on the orientation angle $\psi$, and is not sensitive to the handedness of the excitation, due to the nonchiral geometry (see Suppl. Inf. 4).

\subsubsection{Nonlinear response under linearly polarized excitation}

From analytical considerations (see Appendix~\ref{app:functional2}) we show that structures with near unity SH-CD under linearly polarized excitation ($\chi=0$) feature dependence of SH signal intensity approximated by the following equation:

\begin{equation} 
   I^{2\omega}\approx a+b\cos(2\psi + \psi_0),
    \label{eq:power_rect_hori}
\end{equation}

Fitting of the normalized calculated dependencies at 1550 nm gives for the coefficients $a$ and $b$ values of $0.507$ and $0.489$, respectively. Given the close values of the coefficients, Equation~\ref{eq:power_rect_hori} can be rewritten as $I^{2\omega}\approx 2b\cos^2(\psi+\psi_0/2)+\varepsilon$, with $\varepsilon = a-b = 0.018$.
Such behavior closely resembles the conventional dependence of light intensity transmitted through a rotated linear polarizer \cite{deCeglia:15, Collett1992}. We should note that while working in the region with near unity SH-CD approximately gives the $\cos^2(\psi)$ law, detuning from the optimized wavelength results in decreased SH-CD, changing the shape of the $\psi$ dependence (see Figure~\ref{fig:psi_fit}). 
We find that in the general case for our structure (including cases with the SH response of the same order for RCP and LCP excitation) the fitting function includes an additional angular harmonic:
\begin{align}
    I^{2\omega} = a+b\cos(2\psi+\psi_b) +d\cos(4\psi+\psi_d).
    \label{eq:fit_expr}
\end{align}
where the relative weight of the coefficient $d$ increases with detuning from the optimized wavelength. 
The results of the fit with Equation~\ref{eq:fit_expr} are in perfect agreement with the numerical results, as demonstrated in Figure~\ref{fig:psi_fit}.

\subsubsection{Nonlinear response under circularly polarized excitation}
Next, we focus on the description of SH response under circularly polarized excitation. 
Periodic arrays of AlGaAs strips exhibit $C_{2v}$ symmetry, for which the nonlinear polarization can be expressed in cylindrical coordinates $(r, \varphi, z)$ as follows~\cite{nikitina2024achiral, Nikitina2023Jan}:
\begin{align}
    \nonumber
       \mathbf{P}^{2\omega}_{m_\text{in}}(r,\varphi, z) \propto \sum_\nu e^{\iu 2\varphi\nu}\left( \mathbf{P}^{2\omega}_{-2+2m_\text{in},\nu}(r,z)e^{\pm(-2+2m_\text{in})\iu \varphi}e^{\pm 2\iu \beta} + \right.\\
       \left.+ \mathbf{P}_{2+2m_\text{in},\nu}^{2 \omega}(r,z) e^{\pm(2+2m_\text{in})\iu \varphi} e^{\mp 2 \iu \beta} \right)
       \label{eq:shP_gen}
\end{align}
where summation is over $\nu\in \mathbb{Z}$, $\beta$ is the angle of the crystalline lattice rotation with respect to a vertical mirror plane, and $m_\text{in}$ is the total angular momentum (TAM) projection of the incident wave on the propagation $z$-axis ($m_{\text{in}}= -1$ and $+1$ for RCP and LCP, respectively). The dependence of each term on $\varphi$ corresponds to a TAM projection of the eigenmode, excited by this term. 
Basically, in $C_{2v}$ structures there are only 4 types of eigenmodes: two of them have all possible even values of $m$ and the other two --- odd~\cite{Gladyshev2020Aug}. 
In our considerations, only eigenmodes with even $m$ are excited by all $\mathbf{P}^{2\omega}_{m,\nu}$ terms.
For orthogonal inputs ($m_\text{in}=\pm1$) we can rewrite Equation~\ref{eq:shP_gen} as follows:
\begin{align}
       \mathbf{P}^{2\omega}_{\text{L}}(r, \varphi, z) \propto \sum_\nu e^{2 \iu \varphi\nu} \left( \mathbf{P}^{2\omega}_{0,\nu}(r,z) + \mathbf{P}_{4,\nu}^{2\omega}(r,z) e^{4 \iu\varphi} e^{-4\iu\beta}\right) \nonumber\\
       \mathbf{P}^{2\omega}_{\text{R}}(r, \varphi, z)  \propto \sum_\nu e^{2\iu\varphi\nu} \left( \mathbf{P}^{2\omega}_{0,\nu}(r,z)+\mathbf{P}_{-4,\nu}^{2\omega}(r,z) e^{-4\iu\varphi} e^{4\iu\beta}\right)
\end{align} 
Let us keep in mind that each polarization term then excites eigenmodes of the nanostructure of the corresponding symmetry.
In order to qualitatively analyze the nonlinear CD, we can truncate the series to $\nu = 0$ terms, assuming that they provide the dominant contribution to the induced polarization.
This approximation makes considerations simpler, and does not change the qualitative result.
One can easily extend the analysis to higher-order terms, but the conditions for the dichroism will remain the same.
   
\begin{align}
    \mathbf{P}^{2\omega}_{\text{L}} = \mathbf{P}_{0,0} + \left(\mathbf{P}_{4,0} e^{4\iu\varphi}\right) e^{-4\iu\beta} \nonumber\\
    \mathbf{P}^{2\omega}_{\text{R}} = \mathbf{P}_{0,0}+ \left(\mathbf{P}_{-4,0}e^{-4\iu\varphi}\right) e^{4\iu\beta}
    \label{eq:pol_cros}
\end{align}
We note that, as the periodic structure is not chiral, modes excited by the polarization terms with $|\mathbf{P}_{-4,0}^{2 \omega}|$ and $|\mathbf{P}_{4,0}^{2 \omega}|$ with opposite TAM projections must have the same magnitude. 
In general, two terms excite eigenmodes of the same symmetry, but with different amplitudes. 
One can assume that $|\mathbf{P}_{0,0}^{2 \omega}|$ excites one eigenmode, and  $|\mathbf{P}_{\pm 4,0}^{2 \omega}|$ another one, with different phase and amplitude.
The phase difference between the modes excited by 
$m=0$ term and 
terms with $m =\pm4$ equals $\delta_\text{m}$. 
Here, we also assumed that each term excites mostly one mode, while in reality both terms excite an infinite number of eigenmodes with different amplitudes. 
Also, for tilt angles $\beta \neq \frac{\pi \nu}{4}=0,\pm45^{\circ},\pm90^{\circ},...$, the exponential factors in Equation~\ref{eq:pol_cros} do not coincide $e^{\iu\delta_\text{m}-4\iu\beta} \neq e^{\iu\delta_\text{m}+4\iu\beta}$. 
Therefore, we see that the magnitude of the total radiated power differs between LCP and RCP,
enabling the chiral discrimination and the access to the $S_3$ component.

Also assuming the same absolute values of the amplitudes of the excited eigenmodes and integrating for the SH intensity~\cite{nikitina2024achiral} we obtain:
\begin{align}
    I^{2\omega}\propto (1+\cos(\delta_\text{m} \mp 4\beta))^2
\end{align}
where $\delta_\text{m}$ is the relative phase between the modes, determining the modulation depth of the SH-CD. 
Figure~\ref{fig:scatts_expl}b (left panel) plots the total power as a function of the ellipticity angle. It can be seen that for $U_R$ units as the angle varies from $-\pi/4$ (LCP) to $\pi/4$ (RCP), the intensity of the SH signal gradually increases from approximately zero to its nominal value.

\section{Stokes parameters retrieval and performance estimation}

\begin{figure}[!t]
  \centering\includegraphics[width=\textwidth]{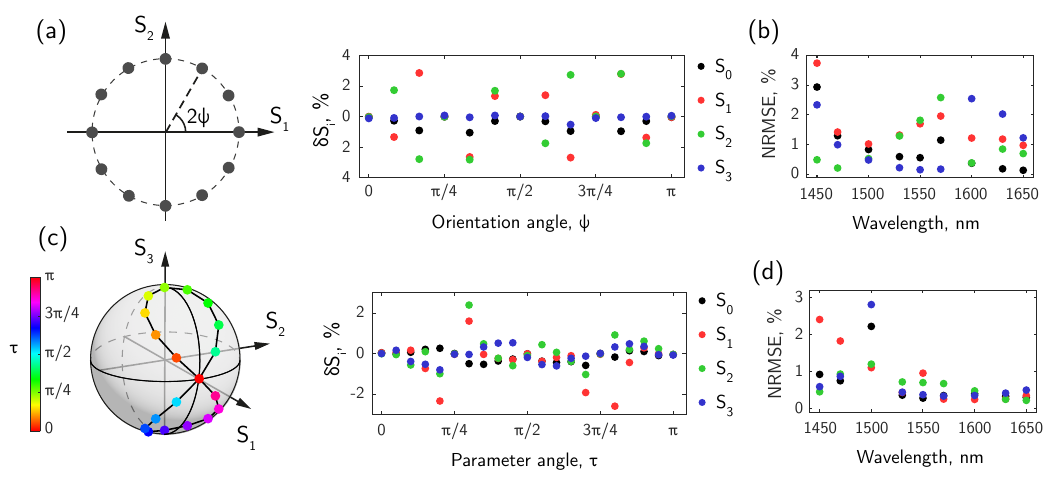}
  \caption{ Stokes parameters retrieval from the second harmonic signal intensity measurements. (a) Reference input of the tested linear SoPs displayed on the Poincar\'e sphere (left panel) and the error of the retrieved Stokes parameters at input wavelength of 1550 nm (right panel). (b) Dependence of the normalized RMSE on the wavelength of linearly polarized input light. (c) Polarization states produced by a quarter-wave plate rotated by an angle $\tau$ (left panel), and the corresponding retrieval error (right panel). (d) Dependence of the normalized RMSE on the wavelength of the input light.   }
  \label{fig3}
\end{figure}

After describing the properties of the nonlinear response of the individual units, we will now discuss how the distinct SH dependencies of a polarimetric super-pixel (or a set of four dielectric polarimetric units) allow to connect the intensities of the SH signal with the polarization state of the input light. First, we consider the procedure allowing for the retrieval of the polarization state from measured SH power.
Next, we show that the response of the units allows for full Stokes polarimetry. The retrieval capabilities of the proposed method are reported in Figure~\ref{fig3}, where the representation of the SoP under analysis is described as a point located on the Poincar\'e sphere~\cite{BornWolf}.

\subsection{Procedure for polarization state retrieval}

To retrieve the Stokes parameters of the input signal, we create a set of lookup tables (LUTs), $T_i(\psi,\chi)$, of size $21\times 11$ $(\psi\times \chi)$, containing the SH power for the corresponding polarization states with fully polarized input. Values between the calculated points are evaluated using linear interpolation. A LUT is calculated for each polarimetric unit at specified input signal intensities and wavelengths. Given the quadratic dependence of the generated SH signal, we assume that the value of the measured SH power, $p_i$, can be obtained by scaling the values taken from the LUTs:
\begin{equation}
    p'_i = \left(\frac{I}{I_c} \right)^2 T_i(\psi, \chi)
\end{equation}
where $p'_i$ is the expected SH power from $i$-th structure, $I$ is the input intensity, $I_c$ is the intensity at which the LUT $(T_i)$ was measured.
With this assumption, we optimize the polarization ellipse angles and the input intensity by minimizing the difference between the scaled LUT values and the measured SH power from each element:
\begin{align}
    \min_{I > 0,\, \psi,\, \chi} 
\left\| 
\left( \mathbf{p} - \alpha\, \mathbf{T}(\psi, \chi) \right) 
\right\|^2, 
\qquad 
\alpha = \left( \frac{I}{I_c} \right)^2, \\
\text{with} \quad \mathbf{p} = [p_1, p_2, p_3, p_4], \quad \mathbf{T} = [T_1, T_2, T_3, T_4].\nonumber
\end{align}
The minimization was performed using a simulated annealing algorithm. 
In order to estimate the accuracy of the retrieval we introduce the retrieval error defined as follows:
\begin{align}
    \delta S_i = \frac{S_i^\text{fit} - S_i^\text{in}}{S_0^\text{in}}
\end{align}
where $S_i^\text{fit}$ is the $i$-th Stokes parameter obtained during the optimization procedure, and $S_i^\text{in}$ is the parameter corresponding to the SoP of the input.
In order to estimate average error for a set of points obtained with the same input intensity $I_\text{in}^{\omega} = S_0^\text{in}$, we use the normalized root mean square error (NRMSE) defined as:
\begin{align}
    \text{NRMSE}_i = \frac{1}{S_0^\text{in}}\sqrt{\frac{1}{N}\sum_{n=1}^N\left(S_i^\text{fit}(n) - S_i^\text{in}(n)\right)^2 }
\end{align}

\subsection{Linear polarimetry}
Figure \ref{fig3}a illustrates the set of SoPs for linearly polarized input, represented by points on the equator of the Poincar\'e sphere ($\chi=0$), together with the error of the retrieved Stokes parameters at an input wavelength of 1550 nm. 
Here during optimization we did not enforce $\chi=0$, allowing the algorithm to retrieve all Stokes parameters. We observe that in this case the absolute error is less than $4\%$. Notably, for points with $\psi$ equal to integer multipliers of $\pi/4$, the error drops to near zero values compared to neighboring points. This can be attributed to the overlap between the points from the test set and the values from the LUTs, whereas the neighboring test points correspond to linearly interpolated LUT values. Figure \ref{fig3}b shows the normalized RMSE a s function of the input wavelength, showing that in the whole range of wavelengths the NRMSE does not exceed $4 \%$.

\subsection{Full Stokes polarimetry}

Next, we focus on full Stokes polarimetry another test set of polarization states, which can be produced experimentally by passing a linearly polarized beam through a rotating quarter-wave plate (QWP) (see Suppl. Inf. 5). In this case the trajectory of the input SoP forms on the sphere a lemniscate like shape (see Figure~\ref{fig3}c, left panel) parametrized by the QWP rotation angle $\tau$. When the fast axis of QWP is aligned with the input polarization ($\tau = 0$), the input state remains unchanged, while at $\tau = \pi/4$, the waveplate transforms the linear input into circularly polarized one. The right panel of Figure~\ref{fig3}c presents the error of the retrieved Stokes parameters at 1550 nm, showing that the absolute error does not exceed 2$\%$. The normalized RMSE for retrieval in the wavelength range 1450–1650 nm remains below 3$\%$, as shown in Figure~\ref{fig3}d.

\subsection{Performance estimation}
Finally, regarding the realization of the proposed device, we provide an estimation for the threshold value of the input power at fundamental frequency required for the operation (see Suppl. Inf.~3.2 for more details). For this purpose, we consider a CCD camera featuring a maximum well capacity of 23000 electrons, an efficiency of $\epsilon\sim 40\%$ at $\lambda^{2\omega}=775$~nm (which corresponds to the SH of a FF beam at 1550~nm), and a pixel size of $p\sim 5$~$\mu$m (\textit{e.g.}, Thorlabs CS2100-USB). Moreover, we consider the polarimetric unit consisting in a grating structure covering a square region with area $A_u=LN\Lambda$, where $L$ is the bar length and $N$ the number of bars within the polarimetric unit.
Given a value of $\xi^{2\omega}\simeq4 \times 10^{-9}$~W$^{-1}$ for $L=25$~$\mu$m and $N=19$, in order to excite $N_e=2000$ electrons in a time interval $\Delta t=5$~s in the case of a single pixel, the required intensity of the pulsed FF excitation can be estimated as 
\begin{equation}
    I_m^{\omega} = I_0^{\omega} \sqrt{\frac{\text{RR}\tau \, 2\pi \hbar c   N_e }{\tilde P^{2\omega}_0LN\epsilon \, \Delta t \, \lambda_{\text{SH}}}} = 23 \ \mbox{mW}/{\mbox{cm}}^2,
\end{equation}
where the pulse duration $\tau=300$~fs and the repetition rate $RR=1$~MHz. We underline that this estimation is based on the assumption that each single analyzer pixel (comprising one polarimetric unit) is imaged onto one single detector pixel.

\section{Conclusions}

In this work, we presented a novel approach to polarimetric imaging based on second-harmonic generation (SHG) in nonlinear flat optics platform. By exploiting the polarization dependence of the nonlinear response in $\chi^{(2)}$ nonlinear crystal of AlGaAs, and by optimizing the geometrical parameters of dielectric gratings, we demonstrated that the intensity of the generated SH signal encodes the full state of polarization of the FF incident light. We analytically analyzed the observed dependencies, revealing the relation between the absolute value of SH dichroism and the shape of the response to the polarization orientation angle.\\

Through numerical simulations, we showed that four identical elements with simple geometry, can serve as micropolarizers of both linear and circular polarizations, selectively enhancing or suppressing SH signals based on the input polarization state. By assembling four such elements into super-pixel arrays, we proposed a complete polarization analyzer capable of retrieving all four Stokes parameters from a single image capture of the SH intensity distribution. We find that such a device can operate with high accuracy (error $<4\%$) in a wide range of wavelengths spanning from 1450 to 1650 nm.\\

Our findings indicate that nonlinear harmonic generation offers a promising route to compact, passive, and cost-effective polarimetric imaging systems.


\section{Acknowledgements}
The authors are thankful to Rinaldo Colombo, Mercedeh Khajavikhan, Giuseppe Leo, Michele Midrio, Luca Palmieri, and Luca Schenato for fruitful discussions.
This work was partially supported by the European Union under the Italian National Recovery and Resilience Plan (NRRP) of NextGenerationEU, of partnership on “Telecommunications of the Future” (PE00000001 - program “RESTART”), Cascade project PRISM - CUP: C79J24000190004, Cascade project SMART - CUP: E63C22002040007, Smart Metasurfaces Advancing Radio Technology (SMART), 
PRIN 2020 project METEOR (2020EY2LJT), METAFAST project that received funding
from the European Union Horizon 2020 Research and Innovation programme
under Grant Agreement No. 899673. A. T. acknowledges the financial support from the University of Palermo through "Fondo Finalizzato alla Ricerca di Ateneo 2025 (FFR2025)". K.F. gratefully acknowledges support from the Alexander von Humboldt Foundation.


%
\setcounter{equation}{0}\renewcommand\theequation{A\arabic{equation}}
\setcounter{figure}{0}\renewcommand\thefigure{A\arabic{figure}}
\setcounter{table}{0}\renewcommand\thetable{A\arabic{table}}

\section{Appendix}
\subsection{Functional dependence on the incident polarization}
\label{app:functional}
In this section, we provide an analytical description of our system, offering valuable insights into the underlying physics. 
Let us denote incident left(right)-circularly polarized plane wave as $\ket{L}(\ket{R})$, or more generally, a beam with TAM projection $m_{\text{in}}= -1(+1)$.
In this basis, the SoP of an arbitrarily polarized input wave $\ket{E^\omega}$ can then be expressed as
\begin{align}
    \ket{E^\omega} \propto \left( \cos\frac{\theta}{2} e^{\iu\psi}\ket{R} + \sin\frac{\theta}{2} e^{-\iu\psi}\ket{L} \right) \sim
    \left(\cos\frac{\theta}{2} \ket{R} + \sin\frac{\theta}{2} e^{-2\iu\psi}\ket{L} \right)  
\end{align}
The equivalence sign $\sim$ is used, because for these two formulae describe the same polarization state, as long as we are not interested in the total phase, but still care about the amplitude, to compare the SH responses to different polarizations. 
We can then describe the input for the SHG process as the tensor product of the incident wave~\cite{Freter2024Oct, nikitina2024achiral, Koshelev2024Apr}, \textit{i.e.} $\ket{E^\omega}\otimes \ket{E^\omega}$.
Let us denote the whole process of SHG by $\mathcal{F}$, because, in this consideration, we are not interested in particular properties of this process. 
The second harmonic field $\vb E^{2\omega}(\vb r)$ can then be  written as 
\begin{align}
\label{eq:tensprod}
    \vb E^{2\omega}(\vb r) = \mathcal{F}(\vb r) (\ket{E}\otimes \ket{E}) = \nonumber \\ =
    \mathcal{F}(\vb r) \left(\left(\cos\frac{\theta}{2}\ket{R} + \sin\frac{\theta}{2} e^{-2\iu\psi}\ket{L} \right) \otimes \left(\cos\frac{\theta}{2}\ket{R} + \sin\frac{\theta}{2} e^{-2\iu\psi}\ket{L} \right) \right) =
    \\ = \mathcal{F}(\vb r) \left(\cos^2\frac{\theta}{2}\ket{R} \otimes \ket{R} + \cos\frac{\theta}{2}\sin\frac{\theta}{2} e^{-2\iu\psi}(\ket{L} \otimes \ket{R} + \ket{R} \otimes \ket{L}) + \sin^2\frac{\theta}{2} e^{-4\iu\psi}\ket{L} \otimes \ket{L}  \right) \nonumber 
\end{align}
The question arises: is $\mathcal{F}(\vb r)$ linear?
First, the fields undergo linear scattering. 
Fields inside the nanostructure do not possess the same shape as for plane wave, but importantly, the scattering process is linear. 
If the incident field is represented as a sum of two waves, the internal field is likewise represented as a sum of the internal fields generated by these two waves.  
We now consider the action of the second-order susceptibility $\chi^{(2)}$, which is nonlinear in the usual sense. 
However, when expressed in the tensor-product formalism as in Equation~\ref{eq:tensprod}, the corresponding operator acts linearly on the tensor-product space. 
The nonlinear polarization then generates the second-harmonic field, which can also be described linearly through Green’s functions.
Specifically, for any input $\ket{E_i}$, the output is expressed as 
$\mathcal{F}(\vb r)(\ket{E_1}\otimes\ket{E_2}+\ket{E_3}\otimes\ket{E_4}) = \mathcal{F}(\vb r)\ket{E_1}\otimes\ket{E_2} + \mathcal{F}(\vb r)\ket{E_3}\otimes\ket{E_4} $.

Let us now introduce the notation for each term in~\eqref{eq:tensprod} of the form:
\begin{align}
    \mathcal{F}(\vb r) \left(\ket{A} \otimes \ket{B}\right) = \vb E^{2\omega}_{AB}.
\end{align}
where $A$ and $B$ take values from the set $\{R,L\}$. Rewriting Equation~\ref{eq:tensprod}, we obtain the expression
\begin{align}
    \vb E^{2\omega} (\vb r) = \left(\cos^2\frac{\theta}{2}\vb E^{2\omega}_{RR} + \cos\frac{\theta}{2}\sin\frac{\theta}{2} e^{-2\iu\psi}(\vb E^{2\omega}_{RL}+\vb E^{2\omega}_{LR}) + \sin^2\frac{\theta}{2} e^{-4\iu\psi}\vb E^{2\omega}_{LL}  \right)
    \label{eq:sh_gen_an}
\end{align}
This is the second-harmonic field at some point $\vb r$.
To obtain the total SH intensity, one must calculate the local intensity by taking the square of the absolute value of the field and then integrate it over space.
The subtle part is that each $\vb E^{2\omega}_{AB}$ is a vector quantity depending on the coordinate, and determining it requires additional calculations.
However, we can look at the symmetry behavior of each of these terms~\cite{nikitina2024achiral}, and immediately determine whether they interfere in the intensity, or not. 
They do interfere, if they possess at least partially coinciding values of total angular momentum projections $m$.
This rule is applicable almost always, except for very specific cases, however, to be sure, one may also consider all the parities~\cite{Frizyuk2019Feb}.
In our case, the nanostructure is of $C_{2v}$ symmetry with GaAs tensor $[001]||z$, so each of the 4 terms generates all possible even $m$ values.
This means that all these partial SH-fields have the same symmetry behavior and interfere. 

\subsection{Dependence of the response with linearly polarized input}
\label{app:functional2}

\begin{figure}[!t]
  \includegraphics[width=1\textwidth]{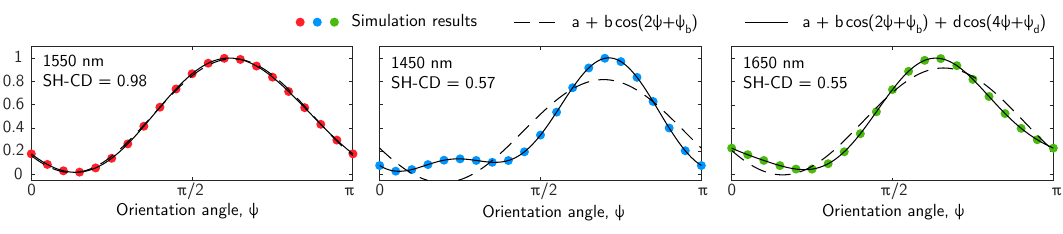}
  \caption{ Dependencies of the normalized SH power generated by the polarimetric unit on the orientation angle for an ellipticity angle $\chi=0$. 
  For small SH-CD values, the angular dependence becomes distorted due to the appearance of an additional harmonic. The simulated data and corresponding fits are shown by markers and lines, respectively. }
  \label{fig:psi_fit}
\end{figure}

Let us now focus on a specific case of linearly polarized input wave ($\theta=\pi/2$) incident on a structure with unity SH circular dichroism.
This means, that one of the responses at SH, $\vb E^{2\omega}_{RR}$ or $\vb E^{2\omega}_{LL}$, is zero. 
Let us consider $\vb E^{2\omega}_{LL}=0$. 
Values of  $\vb E^{2\omega}_{RL}$ and $\vb E^{2\omega}_{LR}$ are not known, but let's assume that they are comparable to $\vb E^{2\omega}_{RR}$.
In this case from Equation~\ref{eq:sh_gen_an} we get the expression in the following form:
\begin{align}
    P^{2\omega}\propto\int |a(\vb r)+e^{-\iu2\psi} b(\vb r)|^2 \dd \vb r = \int \Big(|a(\vb r)|^2+|b(\vb r)|^2+2|a(\vb r)||b(\vb r)|\cos(\delta(\vb r)-2\psi)\Big) \dd \vb r
    \label{eq:cd_un}
\end{align}
and in principle we can assume the phase difference $\delta(\vb r)$ between the fields close to constant (which is quite natural for the far-field of two interfering fields with the same $m$), getting the final expression (as in Equation~\ref{eq:power_rect_hori}) for the dependence on $\psi$:
\begin{align}
    I^{2\omega}\propto\ a+b\cos(2\psi+\psi_b)
    \label{eq:fit_simp}
\end{align}
In the case of imperfect SH circular dichroism (comparable response from both $\vb E_{RR}^{2\omega}$ and $\vb E_{LL}^{2\omega}$ terms), integration of Equation~\ref{eq:sh_gen_an} gives:
\begin{align}
    I^{2\omega}\propto\ a+b\cos(2\psi+\psi_b)+d\cos(4\psi+\psi_d).
    \label{eq:fit_ex}
\end{align}
Figure \ref{fig:psi_fit} shows the numerically calculated dependencies of SH power under linearly polarized excitation on the orientation angle calculated at various wavelengths (colored markers). 
Here we also provide fits of the calculated dependencies with the expressions \eqref{eq:fit_simp} and \eqref{eq:fit_ex}, showed by dashed and solid lines, respectively. We find that the dependence at 1550 nm with SH-CD = 0.98, closely follows Equation~\ref{eq:fit_simp}. In this case, the use of the more accurate expression \eqref{eq:fit_ex} does not significantly improve the accuracy of the fit, which is consistent with the derivation assumptions of near-unity CD. For input wavelengths of 1450 and 1650 nm (with SH-CD of 0.57 and 0.55 respectively), \eqref{eq:fit_simp} fails to describe the observed dependence, resulting in a relatively large error, whereas \eqref{eq:fit_ex} provides an excellent fit. Table~\ref{tb:fit} shows the fitting coefficients for the two expressions.

\begin{table}[h]
\centering
\caption{Fitted parameters for three input wavelengths using simplified and extended formulations.}
\begin{tabular}{llrrrrr}
\hline
$\lambda_{\text{in}},$ nm & Model       & \multicolumn{1}{c}{$a$} & \multicolumn{1}{c}{$b$} & \multicolumn{1}{c}{$\psi_b$, rad.} & \multicolumn{1}{c}{$d$} & \multicolumn{1}{c}{$\psi_d$, rad.} \\
\hline
\multirow{2}{*}{1550} & Simplified & 0.5067 & 0.4892 & 2.3475  &   ---   &   ---  \\
                   & Extended   & 0.5060 & 0.4902 & 2.3497  & 0.01763 & -0.3081 \\
\hline
\multirow{2}{*}{1650} & Simplified & 0.4593 & 0.4597 & 2.1451  &   ---   &   ---  \\
                   & Extended   & 0.4583 & 0.4608 & 2.1488  & 0.09605 & -1.3270 \\
\hline
\multirow{2}{*}{1450} & Simplified & 0.3733 &  0.4431 & 1.9062 &   ---    &   ---  \\
                   & Extended   & 0.3808 &  0.4384 & 1.8742 & 0.18610 & -2.7527 \\
\hline
\end{tabular}
\label{tb:fit}
\end{table}

\setcounter{section}{0}
\setcounter{equation}{0}\renewcommand\theequation{S\arabic{equation}}
\setcounter{figure}{0}\renewcommand\thefigure{S\arabic{figure}}
\setcounter{table}{0}\renewcommand\thetable{S\arabic{table}}

\section*{Supplementary Information}
\section{Simulation of second harmonic generation in periodic structures}


Simulation of the nonlinear response from the gratings was performed using COMSOL Multiphysics\texttrademark \  software in 2D geometry. Linear optical parameters of AlGaAs ($21.9\%$ Al) were taken from Ref.~\cite{papatr2025Jul}, Al$_2$O$_3$ substrate from Ref.~\cite{malitson1972refractive} and the refractive index of the superstrate (air) was set to unity. Nonlinear coefficient $d_{36}$ of AlGaAs was equal to 188 pm/V \cite{ohashi1993determination, Boyd2003}.
In order to mimic the tilt of the gratings with respect to the crystalline lattice, we introduced the rotation in the expression for the nonlinear tensor and the orientation angle $\psi$. For the $U^R_y$ and $U^R_x$ units (upper and lower signs, respectively) the expressions for the polarization are given as follows~\cite{Nikitina2023Jan}:
\begin{align}
P_x &=2\varepsilon_0d_{36}( \pm E_xE_y\pm E_zE_y) \nonumber \\
P_y &=\varepsilon_0d_{36}( \mp E_zE_z\pm E_xE_x \pm2E_zE_x) \nonumber\\
P_z &=2\varepsilon_0d_{36}( \pm E_xE_y \mp E_zE_y)
\label{eq:pol_Run}
\end{align}
and for the $U^L_y$ and $U^L_x$ units (upper and lower signs, respectively):
\begin{align}
P_x &=2\varepsilon_0d_{36}( \mp E_xE_y\pm E_zE_y) \nonumber \\
P_y &=\varepsilon_0d_{36}( \pm E_zE_z\mp E_xE_x \pm2E_zE_x) \nonumber\\
P_z &\propto 2\varepsilon_0d_{36}(\pm E_xE_y \pm E_zE_y)
\end{align}
Here the polarization is defined in a right-handed $xyz$ frame (as used in COMSOL), where the $z$-axis points out of the plane and the $y$-axis is normal to the substrate, directed upward toward the air domain.

\section{Optimization of geometrical sizes}

Optimization was performed using Nelder-Mead algorithm, implemented in COMSOL. We introduce the objective function in the form:
\begin{equation}
    f_{CD} = \left( 1-\frac{I^{2\omega}_\text{RCP} - I^{2\omega}_\text{LCP}}{I^{2\omega}_\text{RCP} + I^{2\omega}_\text{LCP}}\right)^2
\end{equation}
minimization of which maximizes the nonlinear circular dichroism. We optimized the bar width ($w$), period ($\Lambda$) and height ($h$), resulting in  653, 1316 and 246 nm, respectively. We applied constrains on the period of the grating (1250 - 1450 nm), to avoid SH diffraction in large angles and diffraction of FF light in air. In this optimization we obtained SH-CD = 0.9812 and SH-LD = 0.9587. 

In order to determine the effect of linear dichroism optimization, we also introduce corresponding objective function:
\begin{equation}
    f_{LD} = \left( 1-\frac{I^{2\omega}_{||} - I^{2\omega}_\perp}{I^{2\omega}_{||} + I^{2\omega}_\perp}\right)^2
\end{equation}
minimizing the sum $f_{CD}+f_{LD}$. In this case we obtained $w=643$ nm, $\Lambda=1319$ nm and $h=249$ nm, with SH-CD = 0.9790 and SH-LD = 0.9620. Relatively small difference between the two results implies joint requirements on the maximization of CD and LD values.

\section{Performance Estimations}

\subsection{Nonlinear Conversion Coefficient}
The nonlinear conversion coefficient is defined as $\xi^{2\omega}=P^{2\omega}/(P^{\omega})^2$, where $P^{2\omega}$ is the power of the SH radiation generated by the polarimetric unit and $P^{\omega}$ is the power of the FF radiation impinging onto the polarimetric unit. We assume that the polarimetric unit consists in a grating structure covering a square region with area $A_u=NL\Lambda$, where $L$ is the length of the grating bars and $N$ is the number of grating bars. We set $L=25$~$\mu$m, thus $N=L/\Lambda\simeq19$. The quantity $P^{2\omega}$ can be calculated from the results in Figure~2 of the main text, which shows the linear power density ${\tilde{P}}^{2\omega}$ emitted by one unit cell. So, $P^{2\omega}={\tilde{P}}^{2\omega} LN$. On the other hand, regarding the FF radiation, $P^{\omega}=I_0 \cdot A_u=I_0 NL\Lambda$. Therefore, taking ${\tilde{P}}^{2\omega}=3\times10^{-4}$~W/m, the value of the nonlinear conversion coefficient is
\begin{eqnarray}
    \xi^{2\omega}=\frac{P^{2\omega}}{(P^{\omega})^2}=\frac{{\tilde{P}}^{2\omega}}{I_0^2 \Lambda^2 L} \cdot \frac{1}{N}=3.647\times10^{-9} \, {\mbox{W}}^{-1}.
\end{eqnarray}

\subsection{Required Intensity in Practical Realization}

Considering one single pixel of the CCD camera, in order to excite $N_e$ electrons in the time interval $\Delta t$, the required value of the power (average value) of the radiation impinging onto a single pixel is
\begin{eqnarray}
   P_\text{inc,pxl}^{2\omega}=\left( \frac{2\pi\hbar c}{\lambda^{2\omega}} \right) \cdot \frac{N_e}{\epsilon \cdot \Delta t},
\end{eqnarray}
where $E_\text{ph}^{2\omega}=2 \pi \hbar c/\lambda^{2\omega}$ is the energy of a single SH photon and $\epsilon$ is the quantum efficiency of the CCD. Indeed, the quantity $\epsilon \, P_\text{inc,pxl}^{2\omega} \, \Delta t =E_\text{ph}^{2\omega} \, N_e$ describes the amount of energy of SH radiation converted into $N_e$ electrons in one single CCD pixel. With a telescope system the power of the SH radiation reaching one pixel is the same as the power of the SH radiation leaving one polarimetric unit of the analyzer, i.e. $P_\text{inc,pxl}^{2\omega}=P_{\text{out},G}^{2\omega}$. The quantity $P_{\text{out},G}^{2\omega}$ can be related to the power of incident radiation at fundamental frequency $P_{\text{in},G}^{\omega}$ thanks to the nonlinear nonlinear conversion coefficient $\xi^{2\omega}=P^{2\omega}/(P^{2\omega})^2=P_{\text{out},G}^{2\omega}/{\left(P_{\text{in},G}^{\omega}\right)}^2$, which provides $$
P_{\text{in},G}^{\omega}=\sqrt{\frac{P_{out,G}^{2\omega}}{\xi^{2\omega}}}=\frac{1}{\sqrt{\xi^{2\omega}}} \cdot \sqrt{P_\text{inc,pxl}^{2\omega}}.$$
The relation $1/\sqrt{\xi}=I_0 \Lambda \sqrt{LN/{\tilde{P}}_{0}^{2\omega}}$ obtained in Supplementary Sec.~3.1, yields to
$$
P_{\text{in},G}^{\omega}=\frac{I_0}{\sqrt{{\tilde{P}}_{0}^{2\omega}}} \cdot \Lambda \cdot \sqrt{LN} \cdot \sqrt{\left( \frac{2\pi\hbar c}{\lambda^{2\omega}} \right) \cdot \frac{N_e}{\epsilon \cdot \Delta t}},
$$
from which it is possible to evaluate the intensity (mean value) of the radiation impinging onto one polarimetric unit (with area $A_u=NL\Lambda$, as detailed in Supplementary Sec.~3.1):
\begin{eqnarray} \label{eqn:SI_avg_Inte_avg}
I_{\text{in},G}^{\omega}=\frac{P_{\text{in},G}^{\omega}}{A_u}= \sqrt{\left( \frac{I_0^2}{{\tilde{P}}_{0}^{2\omega} } \right) \cdot \left( \frac{2\pi\hbar c}{\lambda^{2\omega}} \right) \cdot \frac{N_e}{\epsilon \cdot \Delta t\cdot L \cdot N}}.    
\end{eqnarray}
At this stage it is worthy to underline that, in Equation~\ref{eqn:SI_avg_Inte_avg}, the terms $I_0$ and ${\tilde{P}}_{0}^{2\omega}$ denote time-averaged quantities (continuous wave, CW, radiation). In the case of instantaneous quantities, as in the case of pulsed (labeled by the subscript $p$) radiation, the following relation should be taken into account: \\$I_{0,p}=I_{0}/\left( \tau_p \cdot RR \right)$ and ${\tilde{P}}_{0,p}^{2\omega}={\tilde{P}}_{0}^{2\omega}/\left( \tau_p \cdot RR \right)$, where $\tau_p$ is the pulse duration and $RR$ is the repetition rate. Therefore, Equation~\ref{eqn:SI_avg_Inte_avg} takes the form
\begin{eqnarray} \label{eqn:SI_avg_Inte_pulsed}
I_{in,G}^{\omega}= \sqrt{\tau_p \cdot RR \cdot \left( \frac{I_{0,p}^2}{{\tilde{P}}_{0,p}^{2\omega} } \right) \cdot \left( \frac{2\pi\hbar c}{\lambda^{2\omega}} \right) \cdot \frac{N_e}{\epsilon \cdot \Delta t\cdot L \cdot N}}.    
\end{eqnarray}

\section{Linear response of polarimetric units}

\begin{figure}[!th]
\centering
\includegraphics[width=0.7\linewidth]{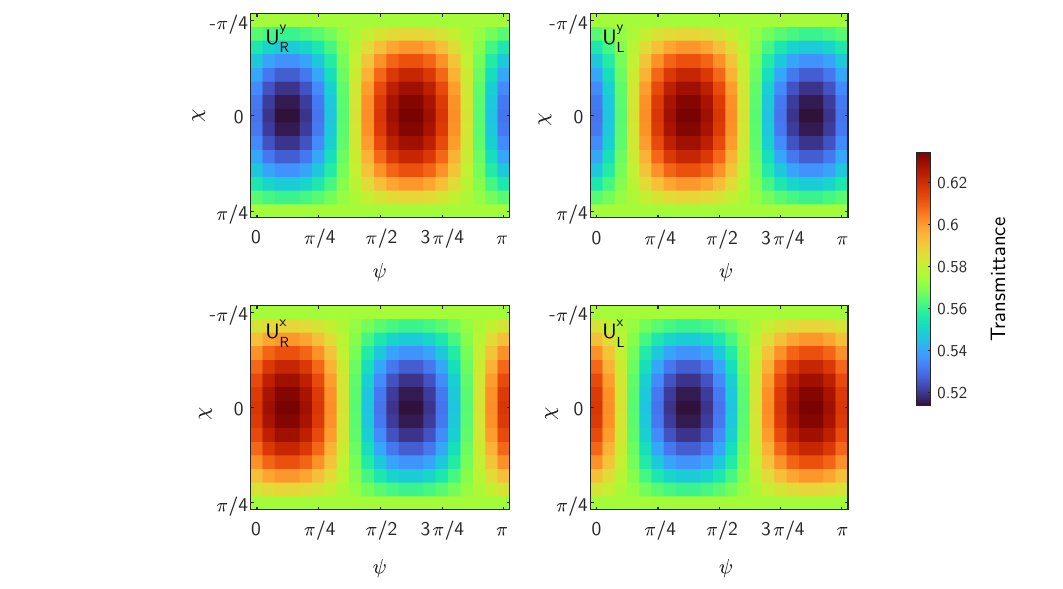}
\caption{ Transmittance of the polarimetric units at 1550 nm. }
\label{lin_resp}
\end{figure}

Figure \ref{lin_resp} shows the transmittance of the four polarimetric units at input irradiation with a wavelength of 1550 nm as a function of the ellipticity $\chi$ and orientation $\psi$ angles. One can observe the modulation of the transmittance of about $5\%$ with the maxima at $\chi = 0$.

\section{Mueller matrices and trajectories on the Poincar\'e sphere}

The Mueller matrix provides a systematic way of representing all of the polarization properties of a sample \cite{polarizedbook}. It is a 4 × 4 matrix that transforms the input Stokes parameters, $S_i$, into the exiting Stokes parameters, $S_i'$, or explicitly:

\begin{equation}
\left(\begin{array}{c}
S_0^{\prime} \\
S_1^{\prime} \\
S_2^{\prime} \\
S_3^{\prime}
\end{array}\right) =\left(\begin{array}{llll}
M_{00} & M_{01} & M_{02} & M_{03} \\
M_{10} & M_{11} & M_{12} & M_{13} \\
M_{20} & M_{21} & M_{22} & M_{23} \\
M_{30} & M_{31} & M_{32} & M_{33}
\end{array}\right)\left(\begin{array}{c}
S_0 \\
S_1 \\
S_2 \\
S_3
\end{array}\right)
\end{equation}

When a polarization element with a Mueller matrix $M$ is rotated about the incident beam propagation direction by an angle $\tau$, the angle of incidence is unchanged. For example, consider a normal-incidence beam passing through an element rotating about its normal, the resulting Mueller matrix $M(\tau)$ can be obtained by applying rotation matrices $R(\tau)$ as follows:

\begin{equation}
\label{rotation}
\begin{aligned}
M(\tau) & =R(\tau) \cdot M \cdot R(-\tau) \\
& =\left(\begin{array}{cccc}
1 & 0 & 0 & 0 \\
0 & \cos 2 \tau & -\sin 2 \tau & 0 \\
0 & \sin 2 \tau & \cos 2 \tau & 0 \\
0 & 0 & 0 & 1
\end{array}\right) \cdot\left(\begin{array}{cccc}
M_{00} & M_{01} & M_{02} & M_{03} \\
M_{10} & M_{11} & M_{12} & M_{13} \\
M_{20} & M_{21} & M_{22} & M_{23} \\
M_{30} & M_{31} & M_{32} & M_{33}
\end{array}\right) \cdot\left(\begin{array}{cccc}
1 & 0 & 0 & 0 \\
0 & \cos 2 \tau & \sin 2 \tau & 0 \\
0 & -\sin 2 \tau & \cos 2 \tau & 0 \\
0 & 0 & 0 & 1
\end{array}\right) .
\end{aligned}
\end{equation}

\begin{figure}[t]
\centering
\includegraphics[width=0.7\linewidth]{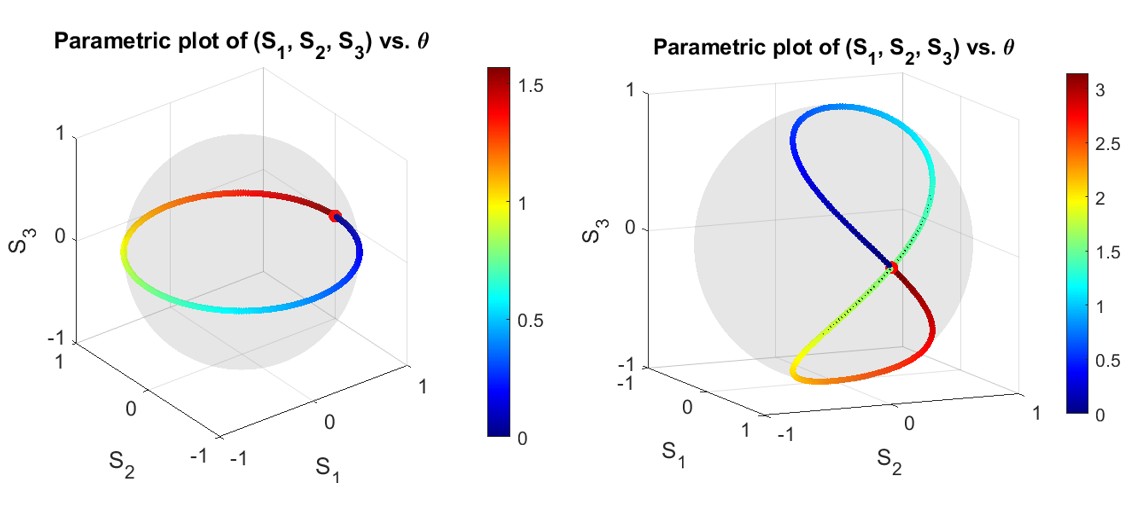}
\caption{Parametric curve of the output Stokes parameters $[S^\prime_0, S^\prime_1,S^\prime_2,S^\prime_3]$ for an input vector $S = [1, 1,0,0]$. Trajectory with a HWR (left) and QWR (right) rotated by an angle $\tau$ on the Poincar\'e sphere. }
\label{8shape}
\end{figure}

A particular type of Mueller matrices are retarders. Retarders are optical elements that introduce different optical path lengths (phases) to two orthogonal eigenpolarization states. A retarder is characterized by the optical path difference between its two orthogonal eigenpolarization states (the retardance $\delta$) and by their corresponding directions, known as the fast axis (smaller optical path length) and the slow axis (larger optical path length). The Mueller matrices for vertical (fast axis aligned along the $y$-axis) quarter wave (QWR) and half wave (HWR) retarders are described as follows:

\begin{equation}
\label{retarders}
\text { QWR }= \quad\left(\begin{array}{cccc}
1 & 0 & 0 & 0 \\
0 & 1 & 0 & 0 \\
0 & 0 & 0 & -1 \\
0 & 0 & 1 & 0
\end{array}\right)
\quad  \quad
\text { HWR } =\quad\left(\begin{array}{cccc}
1 & 0 & 0 & 0 \\
0 & 1 & 0 & 0 \\
0 & 0 & -1 & 0 \\
0 & 0 & 0 & -1
\end{array}\right)
\end{equation}

The variable-angle QWP configuration is a practical method for generating arbitrary polarization states. For example, rotated QWP produces a characteristic “8-shape” trajectory on the Poincar\'e sphere. When the fast axis of the QWP is aligned with the input linear polarization ($\tau = 0$), the polarization state remains unchanged. At $\tau = \pi/4$, the device transforms the linearly polarized input into circular polarization. The corresponding Mueller matrix can be computed using Eqs.~(\ref{rotation}) and (\ref{retarders}), giving for input vector $S = [1,1,0,0]$ the output one that reads:
\begin{align}
    S'_{\text{QWP}} = \left[1, \cos^2(2\tau), -\frac{1}{2}\sin(4\tau), \sin(2\tau)\right]. 
\end{align}

In a similar manner, a different trajectory can be generated by passing a linearly polarized beam through a rotating half-wave plate (HWP). On the Poincar\'e sphere, the action of a HWP corresponds to a $180^\circ$ rotation about the fast axis, which maps linear polarization states to other linear polarization states. As the fast axis of the HWP is rotated, the resulting trajectory forms a great circle on the equatorial plane of the sphere:
\begin{align}
    S'_{\text{HWP}}=[1, \cos(4\tau),\sin(4\tau),0].
\end{align}
When $\tau = 0$, the polarization is aligned with the fast axis and remains unchanged. At $\tau = \pi/4$, the linear polarization is rotated to an orthogonal one.

\section{Electric field distributions in polarimetric unit}

\begin{figure}[h!]
\centering
\includegraphics[width=0.8\linewidth]{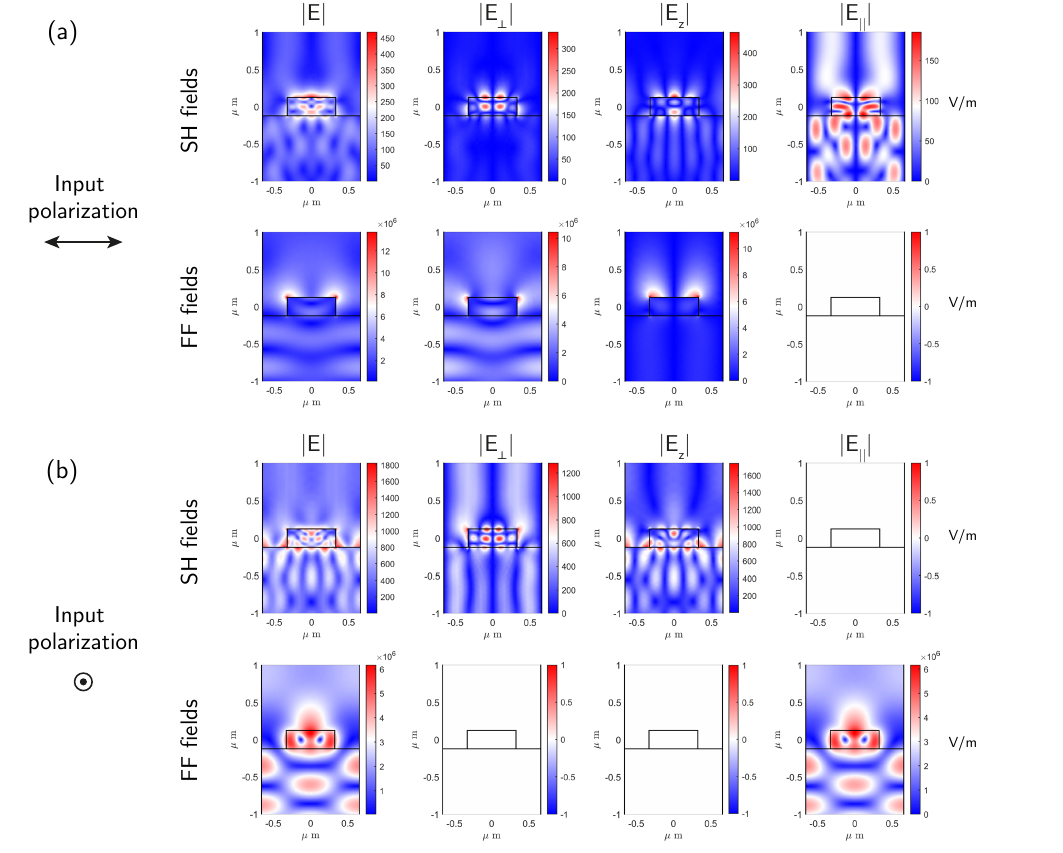}
\caption{(a) Electric field distribution for the input polarized orthogonally to the grating bars, and (b) parallel to the bars.}
\label{fig:fields_lp}
\end{figure}

Figure~\ref{fig:fields_lp} shows the distribution of electric fields at the second harmonic and fundamental frequencies with linearly polarized input light for the $U_y^R$ unit. In the top panel of Figure~\ref{fig:fields_lp}a, the input electric field is polarized horizontally (orthogonally to the grating bars), resulting in excitation of the horizontal ($E_\perp$) and vertical ($E_z$) components, while the near-field component parallel to the bars ($E_{||}$) is zero. At the SH frequency, one can observe excitation of all three components. For input excitation polarized parallel to the grating bars, the only component present at the FF is $E_{||}$ (see Figure~\ref{fig:fields_lp}b). In this case, the SH fields have only horizontal and vertical projections.


\begin{figure}[h!]
\centering
\includegraphics[width=0.8\linewidth]{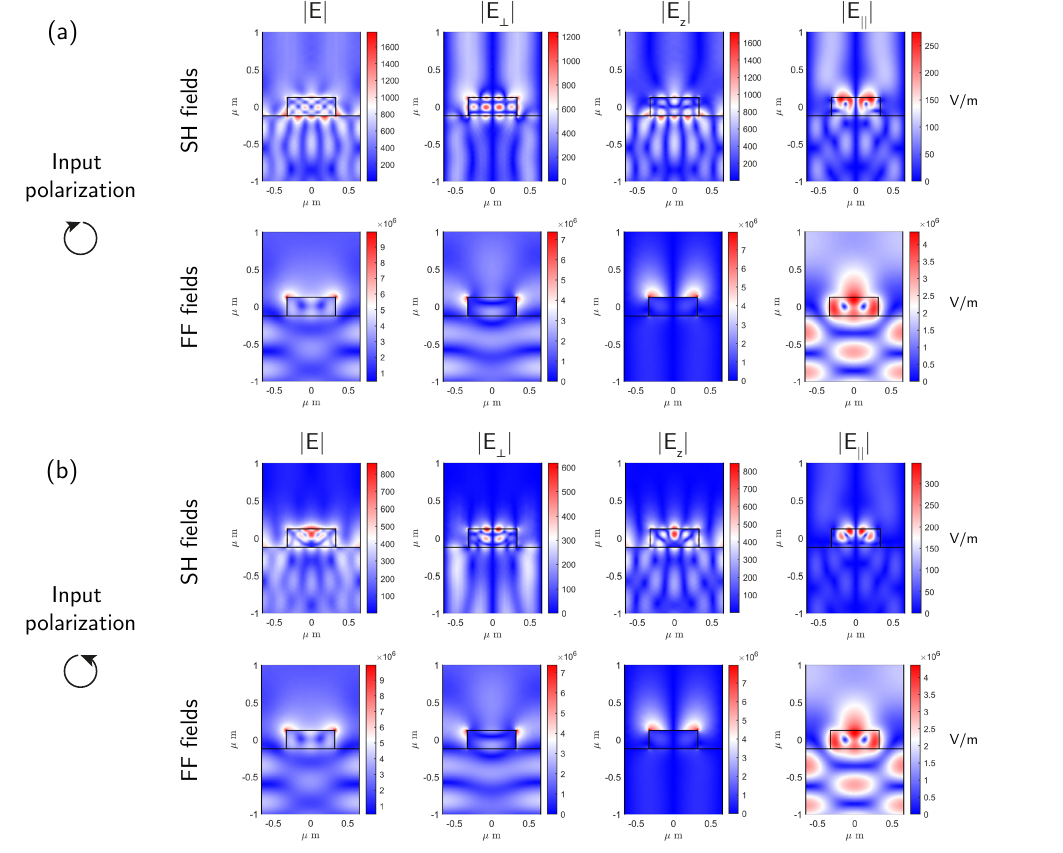}
\caption{(a) Electric field distribution for the RCP ($\chi=\pi/4$) polarized input, and (b) LCP ($\chi=-\pi/4$) polarized input.}
\label{fig:fields_cp}
\end{figure}

In Figure~\ref{fig:fields_cp} are shown electric field distributions for RCP and LCP polarized input. Here, since the structure is not chiral, FF fields distributions are the same for the two cases. 

\newpage

\bibliographystyle{MSP}
\bibliography{bib_polarimeter.bib}

\end{document}